\pdfoutput=1

\documentclass[a4paper,12pt]{article}
\usepackage{jheppub}
\usepackage{amsmath,amssymb,mathrsfs,amsthm,tikz,shuffle,paralist}
\usepackage{dsfont}
\usepackage{color}
\usepackage{enumitem}
\definecolor{darkred}{RGB}{173,34,48}

\usetikzlibrary{snakes}
\usetikzlibrary{calc}
\usetikzlibrary{decorations}

\usepackage{subfigure}
\usepackage{caption}

\newcommand{\dif}{\mathrm{d}} 

\DeclareMathOperator{\Li}{Li} 

\usepackage[utf8]{inputenc}
\usepackage{graphicx}
\usepackage{dcolumn}
\usepackage{bm}

\begin{document}

\title{\Large The Three-loop MHV Octagon from $\bar{Q}$ equations}
\author[a,b]{Zhenjie Li}
\emailAdd{lizhenjie@itp.ac.cn}
\author[c]{Chi Zhang}%
\emailAdd{chi.zhang@nbi.ku.dk}
\affiliation[a]{%
CAS Key Laboratory of Theoretical Physics, Institute of Theoretical Physics, Chinese Academy of Sciences, Beijing 100190, China
}%
\affiliation[b]{%
School of Physical Sciences, University of Chinese Academy of Sciences, No.19A Yuquan Road, Beijing 100049, China
}%
\affiliation[c]{%
Niels Bohr International Academy, Niels Bohr Institute, Copenhagen University, Blegdamsvej 17, 2100 Copenhagen \O{}, Denmark}

\date{\today}

\abstract{The $\bar{Q}$ equations, rooted in the dual superconformal anomalies, are a powerful tool for computing amplitudes in  planar $\mathcal{N}=4$ supersymmetric Yang-Mills theory. By using the $\bar{Q}$ equations, we compute the symbol of the first MHV amplitude with algebraic letters -- the three-loop 8-point amplitude (or the octagon remainder function) -- in this theory. The symbol alphabet for this amplitude consists of 204 independent rational letters and shares the same 18 algebraic letters with the two-loop 8-point NMHV amplitude. 
}

\maketitle

\section{Introduction}
 
Despite the complexity and difficulty of amplitude computations, which grow exponentially as the scattering particles and/or precision (loop order) increase, incredible progress has been achieved in recent decades, especially in planar ${\cal N} {=} 4$ supersymmetric Yang-Mills theory (sYM) due to its considerable symmetries ({\it c.f.} \cite{Arkani-Hamed:2008owk,Roiban:2010kk,Drummond:2010km,Alday:2010kn}). The four- and five-point amplitudes in planar ${\cal N}{=}4$ sYM are captured by the well-known Bern-Dixon-Smirnov (BDS) ansatz~\cite{Bern:2005iz}, as well as the infrared divergences of scattering amplitudes~\cite{Anastasiou:2003kj} for all multiplicities. After subtracting the BDS ansatz, scattering amplitudes with more than 5 particles of the theory are finite functions of cross-ratios, and in particular are expected to be multiple polylogarithms (MPLs)~\cite{Goncharov:2005sla} of weight $2L$ at $L$ loops for MHV and NMHV cases~\cite{ArkaniHamed:2012nw}. Remarkably, the first ``non-trivial'' amplitude -- the 6-point amplitude (or hexagon) -- has been fixed up to seven and six loops for MHV and NMHV cases through a bootstrap program, respectively~\cite{Caron-Huot:2019vjl}, and the 7-point amplitude (or heptagon) has been fixed similarly up to four loops for both cases too~\cite{Dixon:2016nkn,Drummond:2018caf}.

The successes of the hexagon and heptagon programs~\cite{Dixon:2011pw,Dixon:2014xca,Dixon:2014iba,Drummond:2014ffa,Dixon:2015iva,Caron-Huot:2016owq} are based on the following facts: i) each MPL can be characterized by tensor products of logarithmic functions of kinematic variable, \emph{i.e.} the symbol~\cite{Goncharov:2010jf,Duhr:2011zq,Duhr:2012fh}, where various constraints such as the first entry conditions~\cite{Gaiotto:2011dt} and Steinmann relations~\cite{Steinmann1960a,Steinmann1960b} can be easily placed, and ii), more crucially, the sets of the symbol entries, \emph{i.e.} the alphabets, are conjecturally govern by the finite cluster algebras $\rm{Gr}(4,6)$ and $\rm{Gr}(4,7)$ -- finite sets of rational functions of Pl\"{u}cker coordinates. Such a bootstrap program immediately encounter intrinsic obstacles when considering scattering amplitudes with more than 7 external particles: firstly, transcendental functions beyond multiple polylogarithms occur in the scattering amplitudes, such as elliptic polylogarithms~\cite{Bourjaily:2017bsb,Kristensson:2021ani} and beyond~\cite{Bourjaily:2018ycu}, secondly, even for MHV and NMHV cases, we lost control of symbol alphabets from cluster algebras since i) the algebraic letters, which are not rational in the Pl\"{u}cker coordinates anymore, start to appear in two-loop NMHV amplitudes \cite{Zhang:2019vnm}, and ii) the corresponding cluster algebras ${\rm Gr}(4,n)$ with $n>7$ are all of infinite type~\cite{scott2006grassmannians}. On the other hand, the concrete data of scattering amplitudes at high multiplicities are continually necessary: needless to mention their own physical significance, hidden mathematical structures such as positive Grassmannians and cluster algebras~\cite{Arkani-Hamed:2016byb,Golden:2013xva,Drummond:2017ssj,Drummond:2018dfd,Drummond:2019cxm,Arkani-Hamed:2019rds,Henke:2019hve} are revealed by the amplitudes at high multiplicities.

In this paper, we will focus on the three-loop 8-point BDS-subtracted MHV amplitude (which is also \emph{the remainder function} at this loop order) $R_{8,0}^{(3)}$ in planar $\mathcal{N}{=}4$ sYM theory, which is the first multi-loop MHV amplitude containing \emph{algebraic letters}. The appearance of such algebraic letters in three-loop MHV amplitudes is inspired by the previous computations of two-loop NMHV amplitudes for 8 and 9 particles~\cite{Zhang:2019vnm,He:2020vob} and guaranteed by considering external kinematics in two dimensions~\cite{Caron-Huot:2013vda,He:2021fwf}.

To compute this amplitude, we follow refs.~\cite{Zhang:2019vnm,He:2020vob} and use the so-called $\bar{Q}$ equations~\cite{CaronHuot:2011kk}, which are derived from the ``anomalies'' of the \emph{dual} superconformal symmetries of the theory.   
It is well known that the tree-level scattering amplitudes and loop integrands in planar ${\cal N}=4$ sYM theory enjoy both superconformal and dual superconformal symmetries~\cite{Drummond:2006rz,Drummond:2008vq,Korchemsky:2010ut}, which close into an infinite-dimensional Yangian symmetry~\cite{Drummond:2009fd}. At the loop level, the (dual) conformal symmetries are broken by the infrared divergences of loop integrals and restored after subtracting the BDS ansatz. The tree level (classical) dual superconformal symmetries, which are generated by
\begin{equation}
    \bar{Q}_{a}^{A}= \sum_{i=1}^n \chi_{i}^{A}\frac{\partial}{\partial Z_{i}^{a}} ,
\end{equation}
with the momentum twistor $Z_{i}$ and the Grassmann counterpart $\chi_{i}$,
are not preserved in the BDS-subtracted amplitudes yet. In~\cite{CaronHuot:2011kk}, the quantum corrections to the $\bar{Q}$ operator as well as its parity conjugate $Q^{(1)}$ are fixed through the dual Wilson loop picture and the associated Operator Product Expansion (OPE)~\cite{Alday:2010ku}. It follows the $\bar{Q}$ equations and $Q^{(1)}$ equations which are obeyed by the \emph{all loop} BDS-subtracted amplitudes.
Perturbatively, these equations are powerful tools to compute the loop level amplitudes since they express the derivatives of $L$-loop $n$-point amplitudes in terms of one-fold integrals of $(L{-}1)$-loop $(n{+}1)$-point amplitudes. 
In particular, such equations are extensively used in the computation of MHV and NMHV amplitudes where the $\bar{Q}$ equations themselves are sufficient. The $\bar{Q}$ equations have been used to (re-)compute the complete symbol of two-loop MHV for all multiplicities~\cite{CaronHuot:2011ky}, two-loop NMHV heptagon, three-loop MHV hexagon~\cite{CaronHuot:2011kk}, and recently two-loop NMHV for 8 and 9 particles where the algebraic letters appear~\cite{Zhang:2019vnm,He:2020vob}.

As shown in \cite{Zhang:2019vnm}, the main (technical) obstacle in computing MHV/NMHV amplitudes through the $\bar{Q}$ equations is the appearance of algebraic functions in the one-fold integrals mentioned above and overcome by rationalization techniques.
In this paper, we will use the same technique to deal with these algebraic quantities but in symbol integrations. We will see that, as the two-loop 8-point NMHV amplitude inherits its algebraic letters from the four-mass box integrals in the one-loop 9-point N$^{2}$MHV amplitude~\cite{Zhang:2019vnm}, the three-loop 8-point MHV amplitude inherits its algebraic letters from the algebraic words in the two-loop 9-point NMHV amplitude. Moreover, the algebraic letters of the three-loop MHV octagon are the same 18 algebraic letters found in the two-loop NMHV octagon. Since there is no qualitative difference for higher $n$ in this aspect, we expect that algebraic letters appearing in the three-loop MHV amplitudes with more than 8 particles are the same ones in two-loop NMHV amplitudes. For the rational part of the alphabet, we find 24 new letters compared with the 180 rational letters of the two-loop NMHV octagon.

The remainder of the paper is organized as follows. In section \ref{sec:2}, we briefly review amplitudes in planar $\mathcal{N}=4$ sYM theory, the $\bar{Q}$ equations, as well as some technologies of multiple polylogarithms and their symbol. In section \ref{sec:3}, we show how to apply this method to the computation of the three-loop octagon remainder function $R_{8,0}^{(3)}$, in particular the integration of algebraic words through rationalization techniques. In section \ref{sec:4}, we present the symbol alphabet for the three-loop octagon remainder function $R_{8,0}^{(3)}$ and several consistency checks. We conclude and outlook in section \ref{sec:5}.

The explicit expression for the symbol of the three-loop octagon remainder function is included as ancillary files. 
These files are too large to be attached within the arXiv submission, and they are available at \cite{3loopmhvoctagondata}.

\section{Review of $\bar{Q}$ equations and Polylogarithms}\label{sec:2}

In $\mathcal{N}=4$ sYM theory, instead of the usual scattering of particles, we are more interested in the superamplitudes $\mathcal{A}(\Phi_{1},\cdots,\Phi_{n})$ of on-shell superfields
\[
\Phi=g^{+} +\eta^{A}\psi_{A}+\frac{1}{2}\eta^{A}\eta^{B}\phi_{AB} + \frac{1}{3!}\eta^{A}\eta^{B}\eta^{C}\epsilon_{ABCD}\bar{\psi}^{D}
+ \frac{1}{4!}\eta^{A}\eta^{B}\eta^{C}\eta^{D}\epsilon_{ABCD}g^{-}\:,
\]
which are functions of massless momenta $p_{i}$ and Grassmann variables $\eta_{i}$ and encode the super Ward identities. The superamplitude $\mathcal{A}(\Phi_{1},\cdots,\Phi_{n})$ can be naturally decomposed as 
\begin{equation}
    \mathcal{A}(\Phi_{1},\cdots,\Phi_{n})=\mathcal{A}_{n,0}+\cdots+\mathcal{A}_{n,k}+\cdots+\mathcal{A}_{n,n-2}
\end{equation}
due to the R-symmetry $SU(4)$, where $\mathcal{A}_{n,k}$ is a polynomial of degree $4k{+}8$ in $\eta_{i}$ and corresponds to the N$^{k}$MHV sector.

For the planar limit of the theory, the scattering amplitudes have \emph{dual} superconformal symmetries in addition to  usual superconformal symmetries.
Thus, it is convenient to introduce (super) momentum twistor variables~\cite{Hodges:2009hk},
\begin{equation}
    \mathcal{Z}_{i}=(Z_{i}^{a}\vert\chi_{i}^{A}):=(\lambda_{i}^{\alpha},x_{i}^{\alpha\dot{\alpha}}\lambda_{i\alpha}\vert \theta_{i}^{\alpha A}\lambda_{i \alpha})
\end{equation}
with the dual superspace coordinate $(x,\theta)$ defined by $x_{i+1}^{\alpha\dot{\alpha}}-x_{i}^{\alpha\dot{\alpha}}=p_{i}^{\mu}\sigma_{\mu}^{\alpha\dot{\alpha}}=\lambda_{i}^{\alpha}\tilde{\lambda}_{i}^{\dot{\alpha}}$ and $\theta_{i+1}^{\alpha A}-\theta_{i}^{\alpha A}=\lambda_{i}^{\alpha}\eta_{i}^{A}$, which linearly realize the dual superconformal symmetries. In terms of super momentum twistors, we further define the basic $\mathrm{SL}(4)$-invariant $\langle ijkl\rangle:=\epsilon_{abcd}Z_{i}^{a}Z_{j}^{b}Z_{k}^{c}Z_{l}^{d}$ (or the Pl\"{u}cker coordinates of $\mathrm{Gr}(4,n)$), and the basic $R$ invariant~\cite{Drummond:2008vq,Mason:2009qx}
\begin{equation}\label{Rinv}
    [i\,j\,k\,l\,m]:=\frac{\delta^{0\vert 4}(\chi_{i}^{A}\langle jklm\rangle+\text{cyclic})}{\langle ijkl\rangle\langle jklm\rangle
    \langle klmi\rangle\langle lmij\rangle\langle mijk\rangle} \:.
\end{equation}

The infrared divergences and the dual conformal anomalies of planar $\mathcal{N}=4$ sYM amplitudes are captured by the BDS ansatz $\mathcal{A}_{n}^{\text{BDS}}$~\cite{Drummond:2007au}, and \emph{BDS-subtracted amplitudes} $R_{n,k}=\mathcal{A}_{n,k}/\mathcal{A}_{n}^{\text{BDS}}$ are finite and dual conformally invariant (DCI). Furthermore, according to the generalized unitarity method, the $L$-loop BDS-subtracted amplitude $R_{n,k}^{(L)}$ is of the form
\begin{equation}
    R_{n,k}^{(L)}= \sum Y_{n,k}^{\alpha}F_{\alpha}(Z_{i}) \:,
\end{equation}
where leading singularities $Y_{n,k}$'s are Yangian invariants which are independent of loop integrals and fully classified from the positive Grassmannian~\cite{Drummond:2010uq,ArkaniHamed:2009vw,ArkaniHamed:2012nw}, and $F_{\alpha}$ are transcendental functions of cross ratios of Pl\"{u}cker coordinates $\langle ijkl\rangle$ arising from loop integrals. The main interest of this paper is $R_{8,0}^{(3)}$ where $Y_{n,0}$ are simply 1. As we will see below, this amplitude can be computed from one-fold integrals of $R_{9,1}^{(2)}$ where $Y_{n,1}$ are $R$ invariants \eqref{Rinv}.

\subsection{$\bar Q$ equations} 

As shown in~\cite{CaronHuot:2011kk}, the action of the $\bar{Q}$ operator on $R_{n,k}$ is given by an integral over collinear limits of higher-point amplitudes:
\begin{align}
\bar{Q}_{a}^{A} R_{n,k} = \frac{\Gamma_{\rm cusp}}{4}~\operatorname{Res}_{\epsilon=0}\int_{\tau=0}^{\tau=\infty}\Bigl(\mathrm{d}^{2\vert 3}\mathcal{Z}_{n+1}\Bigr)_{a}^{A} 
[ R_{n+1,k+1}
-R_{n,k}R_{n+1,1}^{\text{tree}} ]+ \text{cyclic}  \label{Qbar} \:, 
\end{align}
where $\Gamma_{\rm cusp}$ is the cusp anomalous dimension~\cite{Beisert:2006ez}, and the particle $n{+}1$ is inserted in a collinear limit with $n$ whose (super-) momentum twistor $\mathcal{Z}_{n+1}$ is parametrized by $\epsilon, \tau$:
\begin{equation}
    \mathcal{Z}_{n+1}= \mathcal{Z}_{n}- \epsilon \mathcal{Z}_{n-1} + \frac{\langle n{-}1\,n\,2\,3\rangle}{\langle n\,1\,2\,3\rangle} \epsilon \tau \mathcal{Z}_{1} + \frac{\langle n{-}2\,n{-}1\,n\,1\rangle}{\langle n{-}2\,n{-1}\,2\,1\rangle}\epsilon^{2} \mathcal{Z}_{2} \:.
    \label{colpara}
\end{equation}
The perturbative expansion of eq.\eqref{Qbar} relates $R_{n,k}^{(L)}$ to $ R_{n+1,k+1}^{(L-1)}$ {\it etc.}. 
For the $L$-loop, $n$-point MHV amplitude, the RHS is the $(L-1)$-loop, $(n+1)$-point NMHV amplitude with a tree part. 

The integral measure $(\mathrm{d}^{2\vert 3}\mathcal{Z}_{n+1})_{a}^{A}$ consists of the fermionic part $(\mathrm{d}^{3}\chi_{n+1})^{A}$ and the bosonic part 
\[
    (\mathrm{d}^{2}Z_{n+1})_{a}:=\epsilon_{abcd}Z_{n+1}^{b}\mathrm{d}Z_{n+1}^{c}\mathrm{d}Z_{n+1}^{d} \:.
\]
In the collinear limit eq.(\ref{colpara}), the bosonic measure becomes
\begin{equation}\label{bosonicmeasure}
\frac{\langle n{-}1\,n\,2\,3\rangle}{\langle n\,1\,2\,3\rangle}(\bar{n})_{a}\operatorname{Res}_{\epsilon=0}\epsilon\mathrm{d}\epsilon\int_{0}^{\infty} \mathrm{d}\tau
\end{equation}
with $(\bar{n})_{a}:=(n{-}1\,n\,1)_{a}$. The notation $\operatorname{Res}_{\epsilon=0}$ means to extract the coefficient of $\mathrm{d}\epsilon/\epsilon$ under the collinear limit of $\epsilon\to 0$ where an extra $1/\epsilon^2$ factor comes from the fermionic integrals.

For an MHV or NMHV ($k=0$ or $k=1$) amplitude $R_{n,k}^{(L)}$, once we know
\begin{equation} \label{QbarofR}
    \bar{Q} R_{n,k}^{(L)}=\sum_\alpha Y_{n,k}^{\alpha}F_\alpha\, \bar{Q}\log(a_\alpha) 
\end{equation}
where $F_{\alpha}$ are some transcendental functions and $a_{\alpha}$ are dual conformal invariants, 
then the differential of $R_{n, k}^{(L)}$ can be expressed as\footnote{This total differential is understood to act on transcendental functions $F_{\alpha}$ only.}
\begin{equation} \label{devofR}
    \dif R_{n,k}^{(L)}=\sum_\alpha Y_{n,k}^{\alpha}F_\alpha\, \dif \log(a_\alpha) 
\end{equation}
due to the limited kernel of the operator $\bar Q$ for MHV and NMHV amplitudes~\cite{CaronHuot:2011kk}. For MHV and NMHV cases, $R_{n,k}^{(L)}$ are believed to be polylogarithms of weight $2L$ and hence $F_{\alpha}$ are polylogarithms of weight $(2L{-}1)$, which can be characterized by their \emph{symbols}. Then \eqref{devofR} immediately gives the symbol of $R_{n,k}$ iteratively as ${\cal S}[ R_{n,k}^{(L)}]=\sum_\alpha Y_{n,k}^{\alpha}~{\cal S} [F^{(2L{-}1)}_\alpha] \otimes (a_\alpha)$ where we introduced a superscript $(2L{-}1)$ for $F_{\alpha}$ to indicate their transcendental weights. 
In the next subsection, we will briefly review multiple polylogarithms and their symbols. Currently, let us see how to obtain the RHS of eq.\eqref{QbarofR} from eq.\eqref{Qbar}.

For MHV cases, the RHS of \eqref{Qbar} are integrals of the form
\[
    \operatorname{Res}_{\epsilon=0}\int^{\tau=\infty}_{\tau=0} {\rm d}^{2|3}\mathcal{Z}_{n{+}1}[i\,j\,k\,l\,m]F(\epsilon,\tau)\:,
\]
where $[i\,j\,k\,l\,m]$ is some $R$ invariant and $F$ is a transcendental function. After taking the collinear limit and performing the fermionic integral for the $R$ invariant, one can see that $R$ invariants without $n$ and $n+1$ have zero contribution, and~\cite{CaronHuot:2011kk} 
\begin{align}\label{toqlog2}
    \operatorname{Res}_{\epsilon=0}&\int^{\tau=\infty}_{\tau=0} {\rm d}^{2|3}\mathcal{Z}_{n{+}1}[i\,j\,k\,n\,n{+}1]F(\epsilon,\tau)\nonumber\\
    &=\int^{\infty}_0\biggl({\rm d}\log\frac{\langle Xij\rangle}{\langle Xjk\rangle}\bar Q\log\frac{\langle\bar n j\rangle}{\langle\bar ni\rangle}+{\rm d}\log\frac{\langle Xjk\rangle}{\langle Xik\rangle}\bar Q\log\frac{\langle\bar n k\rangle}{\langle\bar ni\rangle}\biggr)F(\epsilon\to 0,\tau)
\end{align}
for $i,j,k\neq1,n{-}1$, where $X$ is the bi-twistor $Z_{n}\wedge (Z_{n-1}-\frac{\langle n{-}1\,n\,2\,3\rangle}{\langle n\,1\,2\,3\rangle}\tau Z_{1})$, and similarly
\begin{align}\label{toqlog3}
[i\ j\ n{-}1\ n\ n{+1}]&\to {\rm d}\log\frac{\langle Xij\rangle}{\langle X n{-}2 n{-}1\rangle}\bar Q\log \frac{\langle\bar n j\rangle}{\langle\bar ni\rangle},\nonumber\\
[1\ i\ j\ n\ n{+1}]&\to {\rm d}\log\frac{\langle Xij\rangle}{\langle X 12\rangle}\bar Q\log \frac{\langle\bar n j\rangle}{\langle\bar ni\rangle},\\
[1\ i\ n{-}1\ n\ n{+1}]&\to {\rm d}\log\frac{\langle X n{-}2 n{-}1\rangle}{\langle X12\rangle}\bar Q\log \frac{\langle\bar n j\rangle}{\langle\bar ni\rangle}\nonumber
\end{align}
for the boundary cases.

To derive eq.\eqref{QbarofR} from $\bar Q$ equations eq.\eqref{Qbar} for MHV amplitudes, we first need the corresponding NMHV amplitude on the RHS, which can be written as  
\begin{equation*}
   R_{n+1,1}^{(L-1)}= \sum_{1\leq i<j<k<l\leq n} [i\,j\,k\,l\, n+1] F^{(2L-2)}_{n+1,ijkl} \:,
\end{equation*} 
where $F^{(2L-2)}_{n+1,\alpha}$ are transcendental functions of weight $(2L{-}2)$. Then from eq.\eqref{toqlog2} and eq.\eqref{toqlog3}, the $\bar Q R_{n,0}^{(L)}$ can be written as 
\begin{equation}
    \bar Q R_{n,0}^{(L)}= \sum_{i} I^{(2L-1)}_i \bar Q \log \langle \bar n i\rangle+\text{cyclic},
\end{equation} 
where $I_i^{(2L-1)}$ are linear combinations of $\dif\log$ integrals over $\tau$ of collinear limits of $F^{(2L-2)}_{n+1,ijkl}$ and read
\begin{equation} \label{laststepofQbar}
    \sum_\alpha c_{i\alpha} \int_0^\infty \mathrm{d}\log f_\alpha(\tau)~F^{(2L-2)}_{n{+}1,\alpha} (\tau, \epsilon\to 0)
\end{equation}
for some rational coefficients $c_{i\alpha}$.
One may worry about the $\log^{L-1}\epsilon$ divergences arising from the collinear limit, but the divergences are always canceled after integrating over $\tau$, as shown in~\cite{CaronHuot:2011kk}.
Another important feature of $\bar Q$ equations for MHV amplitudes is that the RHS of eq.\eqref{laststepofQbar} is automatically DCI~\cite{CaronHuot:2011kk}: each $I_i$ is DCI, and there are linear relations between these $I_i$'s such that the arguments of $\bar Q\log$ can be collected to be DCI. Therefore, we get the total differential of MHV amplitudes,
\begin{equation}
    \dif R_{n,0}^{(L)}= \sum_{i} I^{(2L-1)}_i \dif \log \langle \bar n i\rangle+\text{cyclic}.
\end{equation} 
The remaining task for us is to calculate these one-fold integrals $I^{(2L-1)}_i$.

The treatment for NMHV cases is similar, the interested readers could consult ref.\cite{He:2020vob} for more details.

\subsection{Multiple polylogarithms, symbols and their integrations}\label{sec:2.2}

The known MHV and NMHV amplitudes in planar $\mathcal N=4$ sYM theory are all described by multiple polylogarithms, including two-loop NMHV amplitudes. Therefore, computing integrals in \eqref{laststepofQbar} is essentially to deal with integrals of multiple polylogarithms. In this subsection, we will briefly review some basic facts about multiple polylogarithms.

Multiple polylogarithms are generalizations of classical polylogarithms, such as $\log,\operatorname{Li}_2$, $\dots$, which are defined by~\cite{goncharov2005galois}
\begin{equation}
    G(a_{1},\ldots,a_{n};z):=\int_{0}^{z}\dif \log(t-a_{1})\, G(a_{2},\ldots,a_{n};t), \label{Gpolylot} 
\end{equation} 
with the starting point $G(;z):=1$ and exceptional cases
\[
    G(\underbrace{0, \ldots, 0}_{k} ; z):=\frac{1}{k !}(\log z)^{k},
\]
where $n$ is called the weight of $G(a_{1},\ldots,a_{n};z)$. For example,
\[
    G(0;x)=\log x,\quad G(1;x)=\log(1-x),\quad G(0,1;x)=-\operatorname{Li}_2(x),\quad \dots.
\]
It is straightforward to see that the total differential of \eqref{Gpolylot} satisfies
\begin{equation}
    \dif G(a_1,\dots,a_n;z) = \sum_{i=1}^n G(a_1,\dots,\hat{a_i},\dots,a_n;z)\,\dif \log \frac{a_i-a_{i{-}1}}{a_i-a_{i{+}1}},
\end{equation}
where $a_i$ is deleted in the $i$-th summand with boundary cases $a_0:=z$ and $a_{n{+}1}:=0$. 

There is a well-known Hopf algebra structure~\cite{Goncharov:2005sla}, which has led to the notion of \textit{symbol}~\cite{Goncharov:2010jf, Duhr:2011zq}. For any weight-$n$ multiple polylogarithm $G^{(n)}$ whose differential reads
\[
\dif G^{(n)}=\sum_i G^{(n-1)}_i \dif\log x_i,
\]
where $\{G^{(n-1)}_i\}$ are polylogarithms of lower weight $(n{-}1)$, its symbol $\mathcal{S}(G^{(n)})$ is recursively defined by
\[
\mathcal{S}(G^{(n)}):=\sum_i \mathcal{S}(G^{(n-1)}_i)\otimes x_i.
\]
For example, the symbol of $G(a_1,\dots,a_n;z)$ is 
\[
    \mathcal{S}(G(a_1,\dots,a_n;z)) = \sum_{i=1}^n \mathcal{S}(G(a_1,\dots,\hat{a_i},\dots,a_n;z))\otimes \frac{a_i-a_{i{-}1}}{a_i-a_{i{+}1}},
\]
which leads to special cases
\[
\mathcal{S}(\log(x))=x,\quad 
\mathcal{S}(\operatorname{Li}_2(x))=-\,(1-x)\otimes x.
\]
The entries of a symbol are called its \textit{letters}, and the collection of all letters is called its \textit{alphabet}.

Conversely, we can first construct the space of symbols, which are spanned by
\[
    \sum_I c_I\,a^I_1\otimes \cdots \otimes a^I_n \:,
\]
where $c_I$ are rational numbers and $a^I_i$ are letters. Symbols of functions should satisfy the \textit{integrability conditions}:
\begin{equation}\label{intcondition}
    \sum_I c_I\,a^I_1\otimes \cdots \otimes a^I_{i-1}\otimes \widehat{a^I_{i}}\otimes \widehat{a^I_{i+1}}\otimes \cdots \otimes a^I_n \, \dif \log a^I_{i}\wedge \dif \log a^I_{i+1}=0
\end{equation}
for any $i$. It follows from the fact that $\dif^2=0$ and the recursive definitions of multiple polylogarithms and symbols.

For our calculation, it's more convenient to perform integrations on the symbol level directly based on the following rules~\cite{CaronHuot:2011kk} (a short proof is reviewed in Appendix \ref{appa}): Suppose we have an integral
\[
\int_a^b {\rm d}\log(t+c)\, (F(t)\otimes w(t)),
\]
where $F(t)\otimes w(t)$ is a linear reducible symbol in $t$, {\it i.e.} its entries are products of powers of linear polynomials in $t$, and $w(t)$ is the last entry. The total differential of this integral is the sum of the following two parts:
\begin{compactenum}[\quad (1)]
\item the contribution from endpoints:
\begin{equation}
    {\rm d}\log(t+c)\Bigl(F(t)\otimes w(t)\Bigr)\Bigr|_{t=a}^{t=b} \Rightarrow \Bigl(F(t)\otimes w(t)\otimes (t+c)\Bigl)\Bigr|_{t=a}^{t=b}, \label{symbolint1}
\end{equation}
\item contributions from the last entry: for a term where $w(t)$ is a constant,
\begin{equation}
\left(\int_a^b {\rm d}\log(t+c)\, F(t)\right){\rm d}\log w \Rightarrow\left(\int_a^b {\rm d}\log(t+c)\, F(t)\right)\otimes w, \label{symbolint2}
\end{equation}
and for a term where $w(t)=t+d$,
\begin{equation}
\left(\int_a^b {\rm d}\log \frac{t+c}{t+d}\, F(t)\right){\rm d}\log (c-d)
 \Rightarrow\left(\int_a^b {\rm d}\log \frac{t+c}{t+d}\, F(t)\right)\otimes (c-d). \label{symbolint3}
\end{equation}
\end{compactenum}

\section{The computation of the three-loop octagon remainder function} \label{sec:3} 

Let's focus on the 3-loop MHV octagon ($n=8$). The starting point is the symbol of the two-loop 9-point NMHV amplitude $R_{9,1}^{(2)}$ \cite{He:2020vob}, which can be written as 
\begin{equation*}
   \mathcal S(R_{9,1}^{(2)})= \sum_{1\leq i<j<k<l\leq 8} [i\,j\,k\,l\, 9] \,\mathcal S(F^{(4)}_{9,ijkl}),
\end{equation*} 
then we follow the procedure mentioned in the last section, the remaining task is to calculate the one-fold integrals of some weight-$4$ symbols. However, some letters of integrands contain algebraic letters of $R_{9,1}^{(2)}$, and hence are no longer linear functions of $\tau$. Thus, we cannot directly apply the symbol integration method mentioned in the last section. We overcome this obstacle by rationalizing these square roots such that letters are linear in the new integration variable.

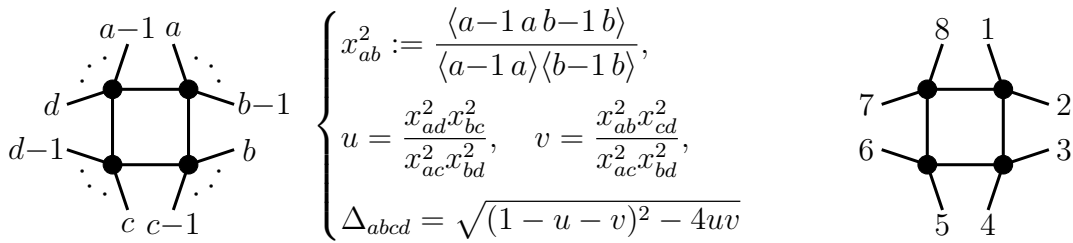
\begin{figure}[t]
    \begin{center}
        \begin{tikzpicture}[baseline={([yshift=-.5ex]current bounding box.center)}]
            \node[fill=black,circle,draw=black, inner sep=0pt,minimum size=7pt] at (-2,0) {};
            \node[fill=black,circle,draw=black, inner sep=0pt,minimum size=7pt] at (-2,-1) {};
            \node[fill=black,circle,draw=black, inner sep=0pt,minimum size=7pt] at (-1,-1) {};
            \node[fill=black,circle,draw=black, inner sep=0pt,minimum size=7pt] at (-1,0) {};
            \draw[line width=0.4mm] (-2,0) -- (-2,-1) -- (-1,-1) -- (-1,0) -- cycle;
            \draw[line width=0.4mm] (-2.6,-0.8) -- (-2,-1);
            \draw[line width=0.4mm] (-2,-1) -- (-1.8,-1.6);
            \draw[line width=0.4mm] (-1,-1) -- (-0.4,-0.8);
            \draw[line width=0.4mm] (-1,-1) -- (-1.2,-1.6);
            \draw[line width=0.4mm] (-2.6,-0.2) -- (-2,0);
            \draw[line width=0.4mm] (-1.8,0.6) -- (-2,0);
            \draw[line width=0.4mm] (-1.2,0.6) -- (-1,0);
            \draw[line width=0.4mm] (-1,0) -- (-0.4,-0.2);
            \node at (-1.2,0.8) {$a$};
            \node at (0,-0.2) {$b{-}1$};
            \node at (-0.2,-0.8) {$b$};
            \node at (-1.2,-1.8) {$c{-}1$};
            \node at (-1.8,-1.8) {$c$};
            \node at (-3,-0.8) {$d{-}1$};
            \node at (-2.8,-0.2) {$d$};
            \node at (-1.8,0.8) {$a{-}1$};
            \node at (-2.4,0.1) {$\cdot$};
            \node at (-2.3,0.3) {$\cdot$};
            \node at (-2.1,0.4) {$\cdot$};
            \node at (-0.9,-1.4) {$\cdot$};
            \node at (-0.7,-1.3) {$\cdot$};
            \node at (-0.6,-1.1) {$\cdot$};
            \node at (-0.6,0.1) {$\cdot$};
            \node at (-0.7,0.3) {$\cdot$};
            \node at (-0.9,0.4) {$\cdot$};
            \node at (-2.4,-1.1) {$\cdot$};
            \node at (-2.3,-1.3) {$\cdot$};
            \node at (-2.1,-1.4) {$\cdot$};
        \end{tikzpicture}
        $\displaystyle
            \begin{cases}\displaystyle
                x_{ab}^2 := \frac{\langle a{-}1\,a\,b{-}1\,b\rangle}{\langle a{-}1\,a\rangle\langle b{-}1\,b\rangle},\\[12pt]
                u=\displaystyle\frac{x_{ad}^{2}x_{bc}^{2}}{x_{ac}^{2}x_{bd}^{2}},\quad v=\displaystyle\frac{x_{ab}^{2}x_{cd}^{2}}{x_{ac}^{2}x_{bd}^{2}},\\[12pt]
                \Delta_{abcd} =\sqrt{(1-u-v)^{2}-4uv}
            \end{cases}
        $
    \qquad 
    \begin{tikzpicture}[baseline={([yshift=-.5ex]current bounding box.center)}]
        \node[fill=black,circle,draw=black, inner sep=0pt,minimum size=7pt] at (-2,0) {};
        \node[fill=black,circle,draw=black, inner sep=0pt,minimum size=7pt] at (-2,-1) {};
        \node[fill=black,circle,draw=black, inner sep=0pt,minimum size=7pt] at (-1,-1) {};
        \node[fill=black,circle,draw=black, inner sep=0pt,minimum size=7pt] at (-1,0) {};
        \draw[line width=0.4mm] (-2,0) -- (-2,-1) -- (-1,-1) -- (-1,0) -- cycle;
        \draw[line width=0.4mm] (-2.6,-0.8) -- (-2,-1);
        \draw[line width=0.4mm] (-2,-1) -- (-1.8,-1.6);
        \draw[line width=0.4mm] (-1,-1) -- (-0.4,-0.8);
        \draw[line width=0.4mm] (-1,-1) -- (-1.2,-1.6);
        \draw[line width=0.4mm] (-2.6,-0.2) -- (-2,0);
        \draw[line width=0.4mm] (-1.8,0.6) -- (-2,0);
        \draw[line width=0.4mm] (-1.2,0.6) -- (-1,0);
        \draw[line width=0.4mm] (-1,0) -- (-0.4,-0.2);
        \node at (-1.2,0.8) {$1$};
        \node at (-0.2,-0.2) {$2$};
        \node at (-0.2,-0.8) {$3$};
        \node at (-1.2,-1.8) {$4$};
        \node at (-1.8,-1.8) {$5$};
        \node at (-2.8,-0.8) {$6$};
        \node at (-2.8,-0.2) {$7$};
        \node at (-1.8,0.8) {$8$};
    \end{tikzpicture}	
    \end{center}
    \caption{\label{fig:wide}A general four-mass box (left) and a octagon one (right).}
\end{figure}

There are nine square roots in $R_{9,1}^{(2)}$, which correspond to nine four-mass boxes (see Fig. \ref{fig:wide}) :
\[
I_{1,3,5,7},I_{1,3,5,8},I_{1,3,6,8},I_{1,4,6,8},I_{2,4,6,8},I_{2,4,6,9},I_{2,4,7,9},I_{2,5,7,9},I_{3,5,7,9},
\]
where the (normalized) four-mass box integral $I_{a,b,c,d}$ is a weight-2 polylogarithm:
\begin{align}
    I_{a,b,c,d}:=\int \dif^{4} x_{0} \frac{-x_{ac}^{2}x_{bd}^{2}\Delta}{x_{0a}^{2}x_{0b}^{2}x_{0c}^{2}x_{0d}^{2}} =
    -\Li_{2}(z)+\Li_{2}(\bar{z})-\frac{1}{2}\log(z\bar{z}) \log\biggl(\frac{1-z}{1-\bar{z}}\biggr) \label{boxintegral}
\end{align}
with
\begin{align}
    u&:=\frac{x_{ab}^{2}x_{cd}^{2}}{x_{ac}^{2}x_{bd}^{2}}\:,\quad v:=\frac{x_{bc}^{2}x_{da}^{2}}{x_{ac}^{2}x_{bd}^{2}} \:,\quad  \Delta:=\sqrt{(1-u-v)^{2}-4uv} \:, \label{uvdel} \\
    z&=\frac{1}{2}\bigl(1+u-v+\Delta\bigr)\:, \qquad \bar{z}=\frac{1}{2}\bigl(1+u-v-\Delta\bigr)\:,\label{zzbar}
\end{align}
and its symbol is 
\[
    \mathcal S(I_{a,b,c,d})=\frac 12\biggl(v\otimes \frac{z}{\bar z}-u\otimes \frac{1-z}{1-\bar z}\biggr).
\]
The subscript $a,b,c,d$ will be restored to indicate the specific box when necessary, otherwise suppressed. It is clear that only $2$ of them, 
\[ 
   \Delta_{3,5,7,9},\:\:\Delta_{2,4,6,9}\:,
\]
can potentially contribute square roots $\Delta_{1,3,5,7}$ (see Fig.~\ref{fig:wide}) and  $\Delta_{2,4,6,8}$ after taking the collinear limit. 

Now let us discuss how to perform $\tau$ integrals involving square roots at the symbol level. Such a technique has also been used in the calculation of two-loop double pentagon integrals or other integrals involving square roots of four-mass box types from the Wilson loop representation~\cite{He:2020lcu}.

\paragraph*{Rationalization}
The $\tau$-integrals involving square roots are of the type
\[
\int \dif \log(\tau-C_{1})\,I(z(\tau),\bar z(\tau))\otimes \frac{z(\tau)-r(\tau)}{\bar z(\tau)-r(\tau)}\otimes (C_{2}\tau-C_{3}),
\]
where $I$ is the four-mass box function, $r(\tau)$ are rational in $\tau$, while $C_{1},C_{2},C_{3}$ are simply constants. We can first apply the symbol integration rules \eqref{symbolint1} -- \eqref{symbolint3} once and conclude that the symbol of the above integral should be the sum of boundary terms and a linear combination of the following terms
\begin{equation}\label{integrand1}
\left(\int \dif\log(\tau-\tilde{C})\,I(z(\tau),\bar z(\tau))\otimes \frac{z(\tau)-r(\tau)}{\bar z(\tau)-r(\tau)}\right)\otimes  f(\tau),
\end{equation}
where $f$ is a rational function of $\tau$, and $\tilde{C}=C_{1}$ or $C_{3}/C_{2}$. However, such a procedure cannot be repeated for the integral of this weight-3 symbol due to the appearance of square roots (which are implicitly contained in $z(\tau)$ and $\bar{z}(\tau)$).

We need to rationalize these square roots by a change of variable. There is a subtlety caused by the fact that the integrand in \eqref{integrand1} itself is not integrable without the support of the other part of the two-loop 9-point NMHV amplitude. Making the variable substitution only for \eqref{integrand1} and then applying the symbol integration algorithm \eqref{symbolint1} -- \eqref{symbolint3} naively will miss some terms. We will return to this point later. For now, let us first give the variable substitution rationalizing the integrand of the form \eqref{integrand1}.

Suppose $z\bar z=u_{a,b,c,d(\tau)}$ and $(1-z)(1-\bar z)=v_{a,b,c,d(\tau)}$, where $d(\tau)$ is a linear combination of two bi-twistors, $d(\tau)=d_0+C\tau d_1$, then there exist constants $p$ and $q$ such that
\begin{equation}
pu_{a,b,c,d(\tau)}+qv_{a,b,c,d(\tau)}=1 \:. \label{pqequation}
\end{equation}
Indeed, one can find the following explicit solution
\[
p=\frac{x_{ac}^2 (x_{bd_1}^2 x_{cd_0}^2- x_{bd_0}^2 x_{cd_1}^2)}{x_{bc}^2(x_{ad_1}^2  x_{cd_0}^2-x_{ad_0}^2  x_{cd_1}^2)},\quad q=\frac{x_{ac}^2(x_{ad_1}^2 x_{bd_0}^2-x_{ad_0}^2 x_{bd_1}^2)}{x_{ab}^2 (x_{ad_1}^2 x_{cd_0}^2-x_{ad_0}^2  x_{cd_1}^2)}.
\]
Moreover, it follows from eq.\eqref{pqequation} that $z$ and $\bar{z}$ are related by a M\"obius transformation 
\[
\bar z=\Lambda(z):=\frac{q z-q+1}{(p +q) z-q} \:,
\]
where $\Lambda^2=\operatorname{id}$. Therefore, $\tau\to \tau(z)=p(z)/q(z)$ with two quadratic polynomials $p$, $q$ in $z$ is a suitable rationalization for our cases, we can express $\bar z$ or any other rational function of $z$, $\bar z$ and $\tau$ as a rational function of $z$. For example, the last entry of the integrand of eq.\eqref{integrand1} becomes
\[
\frac{z(\tau)-r(\tau)}{\bar z(\tau)-r(\tau)}
=\prod_i \left(\frac{z-a_i}{\bar z-a_i}\right)^{n_i},
\]
where $\{a_i\}$ are constants and $n_i$ are rational numbers.

Then we can continue the symbol integration in terms of $z$ for 
\begin{equation}\label{prot2}
    \int_{z(\tau=0)}^{z(\tau=\infty)} \dif \log\left(f(\tau)=c_0 \frac{(z-c) (z-\bar c)}{(z-1)(z-\bar 1)}\right)\,I(z,\bar z)\otimes \frac{z-a}{\bar z-a}, 
\end{equation}
where $a$, $c$ and $c_0$ are constants and a shorthand $\bar x:=\Lambda(x)$ is introduced. Then the symbol integration gives
\begin{align}
&\biggl(\int_{z(\tau=0)}^{z(\tau=\infty)} \dif \log\frac{z - c}{z - \bar c}\: I(z,\bar{z})\biggr)\otimes \frac{a - c}{a - \bar c} - 
I_1\otimes \frac{a-1}{a - \bar 1} +\frac{1}{2} I_{\overline{\infty}}\otimes \frac{(a-c) (a-\bar c)}{(a-\bar 1)(a-\bar{1})}   \nonumber\\
&\qquad \qquad \qquad +
 \Bigl(I\otimes \frac{z-a}{\bar z-a}\otimes f(\tau)\Bigr)\Bigr\vert_{z(\tau=0)}^{z(\tau=\infty)} - I_a\otimes f(\tau_a),\label{prot1}
\end{align}
where $\tau_a$ is defined by $z(\tau=\tau_a)=a$ and 
\[
I_x:=\int_{z(\tau=0)}^{z(\tau=\infty)} \dif \log\frac{z - x}{\bar z - x}\,I(z,\bar z)=\Bigl(I\otimes  \frac{z - x}{\bar z - x}\Bigr)\Bigr\vert_{z(\tau=0)}^{z(\tau=\infty)}+\text{rational terms} \:,
\]
on the support $\bar z=\Lambda(z)$. In our case, the $\dif \log$ measures can always be written in the form eq.\eqref{prot2}. Note that in eq.\eqref{prot1}, besides those terms given by the symbol integration algorithm \eqref{symbolint1} -- \eqref{symbolint3}, we add a term 
\[
    \biggl(\Bigl(I\otimes  \frac{z - a}{\bar z - a}\Bigr)\Bigr\vert_{z(\tau=0)}^{z(\tau=\infty)} - I_a\biggr)\otimes c_0 \:.
\]
This is because the partial derivative of \eqref{prot2} with respect to $c_0$ vanishes in terms of $z$, but it does \emph{not} vanish in terms of $\tau$. The interested readers may refer to Appendix \ref{appa} for more details.



Note that $c$ in \eqref{prot2} could be algebraic and contain a square root which is neither $\Delta_{1,3,5,7}$ nor $\Delta_{2,4,6,8}$. However, such terms from different last entries $(z-a_i)/(\bar z-a_i)$ always cancel:
\[
\left(\int \dif\log\frac{z - c}{z - \bar c} \: I(z,\bar{z})\right)\otimes \biggl(\prod_i\biggl|\frac{a_i - c}{a_i - \bar c}\biggr|^{n_i}=1\biggr)=0,
\]
where $n_i$ is the corresponding rational number factor.

At this stage, one can check that $\tau$-integrals are convergent by regulating the integration region. 
Its $\epsilon\to 0$ limit is also finite which finishes the bosonic integration.

\section{Results and Consistency Checks} \label{sec:4} 

The symbol of the three-loop MHV octagon can be written as $R_{8,0}^{(3)}=\sum_{i=2}^6  F_{i}\otimes \langle \bar 8i\rangle+\text{cyc.}$, these four coefficients are related by reflections
\[
    F_6=F_2|_{i\to 8-i},\quad F_4=F_4|_{i\to 8-i} ,\quad F_5=F_3|_{i\to 8-i}.
\]
The three symbol coefficients $\{F_2,F_3,F_4\}$ are recorded in the files {\tt F2.m}, {\tt F3.m} and {\tt F4.m} (see \cite{3loopmhvoctagondata}). On average, each symbol has around $10^8$ terms.

The symbol alphabet of $R_{8,0}^{(3)}$ consists of $204$ multiplicative-independent rational letters and 18 independent DCI algebraic letters. The 204 rational letters are organized as follows
    \begin{itemize}
        \item $\binom{8}{4}-2=68$ : all $\langle abcd\rangle$ except $\langle 1357\rangle$ and $\langle 2468\rangle$;
        \item 1 cyclic class of $\langle 12(345)\cap (678)\rangle$;
        \item 7 cyclic classes of $\langle 1(ij)(kl)(mn)\rangle$ with $2\leq i<j<k<l<m<n\leq 8$;\\
         5 cyclic classes of $\langle 1(28)(kl)(mn)\rangle$ with $2<k<l<m<n< 8$;
        \item 5 cyclic classes of 
        $
        \langle \bar 2\cap \bar 4\cap (568) \cap \bar 8\rangle,
        \langle \bar 2\cap \bar 4\cap \bar 6 \cap (681)\rangle,
        \langle (127)\cap (235)\cap \bar 5 \cap \bar 7\rangle,
        \langle (127)\cap \bar 3\cap (356) \cap \bar 7\rangle,
        \langle \bar 2\cap (278)\cap (346) \cap \bar 6\rangle.
        $
    \end{itemize}
Here we introduce the notations $\bar a=(a{-}1,a,a{+}1)$ and
\begin{align*}
&\langle a(bc)(de)(fg)\rangle:=\langle abde\rangle \langle acfg\rangle-\langle acde\rangle \langle abfg\rangle,\\
&\langle ab(cde)\cap (fgh)\rangle:=\langle abde\rangle \langle cfgh\rangle+\langle abec\rangle \langle dfgh\rangle+\langle abcd\rangle \langle efgh\rangle,\\
&\langle (a_1b_1c_1)\cap (a_2b_2c_2)\cap (a_3b_3c_3)\cap (a_4b_4c_4)\rangle:=\langle (a_1b_1c_1)\cap (a_2b_2c_2),(a_3b_3c_3)\cap (a_4b_4c_4)\rangle.
\end{align*}
All of these rational letters fall into the predictions from the tropical Grassmannian $\operatorname{Tr}(4,8)$ \cite{Henke:2019hve,Drummond:2019cxm}, but they don't cover the whole prediction, even the smallest dihedrally complete set containing 280 letters. Comparing with the 180 rational letters of $R_{8,1}^{(2)}$~\cite{Zhang:2019vnm}, there are 24 new rational letters: 
\[\text{cyclic images of } \langle 1(23)(46)(78)\rangle,\: \langle \bar 2\cap \bar 4\cap (568) \cap \bar 8\rangle \text{ and } \langle \bar 2\cap \bar 4\cap \bar 6 \cap (681)\rangle.
\]

The 18 algebraic letters are the same as those in the two-loop NMHV octagon. 
The algebraic words of the $R^{(3)}_{8,0}$ can be organized in the form 
\begin{equation}
    I_{1,3,5,7}\otimes \biggl(\chi_{1,3,5,7}(a_1)\otimes R_1+ \chi_{1,3,5,7}(a_2)\otimes R_2+\frac{1}{4}\biggl( \frac{\chi_{1,3,5,7}(0)}{\chi_{1,3,5,7}(1)}\otimes R_3\biggr)\biggr) + \text{cyc.}
\end{equation}
where $\chi_I(a):=(z_I-a)/(\bar z_I-a)$,
\[
a_1=\frac{\langle 4567\rangle \langle 1258\rangle}{\langle 2567\rangle \langle 1458\rangle},\quad
a_2=\frac{\langle 4567\rangle \langle 1358\rangle}{\langle 3567\rangle \langle 1458\rangle}
\]
and $R_1$, $R_2$, $R_3$ are three integrable weight-3 symbols of length $2066$, $5519$, $6392$ respectively. The nine algebraic letters with the square root $\Delta_{1,3,5,7}$ are 
\[
\{\chi_{1,3,5,7}(a_1),\chi_{1,3,5,7}(a_2)\}|_{i\to i+2k} \text{ for $k=0,1,2,3$ and } \frac{\chi_{1,3,5,7}(0)}{\chi_{1,3,5,7}(1)},
\]
where the last letter is invariant under $i\to i+2$, and the other nine algebraic letters with the square root $\Delta_{2,4,6,8}$ are 
\[
\{\chi_{1,3,5,7}(a_1),\chi_{1,3,5,7}(a_2)\}|_{i\to i+2k+1} \text{ for $k=0,1,2,3$ and } \frac{\chi_{2,4,6,8}(0)}{\chi_{2,4,6,8}(1)}.
\]
It's important to notice that algebraic letters can only appear at the second and third entries of the symbol, which is still needed to be understood. The whole algebraic part is recorded in the file {\tt algebraic\_part.m}.

We have performed various consistency checks on our results, for example, cyclicity, dual conformal invariance, and the first entries being of the form $\langle i\,i{+1}\,j\,j{+}1\rangle$; Here we present details for the more non-trivial checks, including integrability and collinear limits.

\paragraph*{Integrability.} It is crucial that the whole data, which is of the form $ \sum_i F_i \dif\log x_i$, can be integrated to a function. This is guaranteed if 
\[
    \sum_i \mathrm{d}\log F_i\wedge \mathrm{d}\log x_i=0.
\]
In terms of the symbol, the result is integrable if and only if the symbol vanishes after taking the last two entries to be the wedge of their $\mathrm{d}\log$'s  (see \cite{Chen:1977oja,brown2009multiple}).
We only need to calculate all two forms
\[
    \dif\log a_i\wedge \dif \log a_j
\]
for all $a_i$ and $a_j$ in the alphabet, which can be numerically done by using any full-rank parametrization of $\mathrm{Gr}(4,8)$. After inserting these numerical two forms, we finally checked that this huge data is integrable. As far as we know, there's no very efficient way to find all linear relations between these two forms by now, which is also crucial for bootstrapping.

\paragraph*{Collinear limits.} We check that the MHV octagon reduces to the MHV heptagon upon taking the collinear limit: first set 
 \begin{equation*}
     Z_8\to Z_7+
       \epsilon \frac{\langle 1257\rangle }{\langle 1256\rangle } Z_6 + 
       \epsilon \tau \frac{\langle 2567\rangle }{\langle 1256\rangle } Z_1 + \eta \frac{\langle 1567\rangle }{\langle 1256\rangle } Z_2
 \end{equation*}
with fixed $\tau$, then take the limit $\eta\to 0$ before $\epsilon\to 0$. After taking such limits and keeping leading terms of $\eta$ and $\epsilon$, it is highly non-trivial that this limit does not depend on the parameters $\eta$, $\epsilon$ and $\tau$, {\it i.e.} it is a smooth limit. Moreover, we find this limit is precisely the three-loop MHV heptagon~\cite{Drummond:2014ffa}.  In particular, all algebraic letters disappear in this limit. 

\section{Conclusion and Outlook} \label{sec:5} 

In this paper, with the two-loop 9-point NMHV amplitude as the input, we computed the symbol of the three-loop octagon remainder function $R_{8,0}^{(3)}$, which is the first three-loop amplitudes containing algebraic letters, by using the $\bar{Q}$ equations. The alphabet of the symbol $\mathcal{S}(R_{8,0}^{(3)})$ consists of 18 algebraic letters and 204 rational ones, where 24 of the rational letters are new compared with the symbol alphabet of the two-loop 8-point NMHV amplitude. Since there is no qualitative difference for higher points, we expect that the three-loop MHV amplitudes share the same algebraic letters with the two-loop NMHV amplitudes for all multiplicities. 

Although there are new letters in the three-loop MHV octagon compared with the known two-loop octagons, the symbol alphabet is covered by the prediction from the tropical Grassmannian $\mathrm{Tr}(4,8)$~\cite{Henke:2019hve,Drummond:2019cxm} which consists of $280$ rational letters and the same 18 algebraic letters. 
It is highly desirable to understand how the alphabets for MHV octagons differ at different loop orders. Furthermore, 
as reported in \cite{He:2021mme}, instead of the usual BDS-subtracted amplitudes, the three-loop MHV octagon with \emph{minimal} subtraction~\cite{Golden:2018gtk} fulfills the (extended) Steinmann relation. There are other arguments~\cite{Henke:2019hve,Herderschee:2021dez} that suggest larger alphabets for octagons, however, it is worthwhile to survey the bootstrap program for octagons already. For example, it would be interesting to see whether the usual constraints used in the bootstrap program are sufficient to fix the known two-loop and three-loop octagons with a redundant symbol alphabet, say that one given by the tropical Grassmannian $\mathrm{Tr}(4,8)$. 

By definition, the $\bar{Q}$ and $Q^{(1)}$ equations are sufficient to fix arbitrary amplitudes in planar $\mathcal{N}=4$ sYM theory up to some Yangian invariants. Thus, a very important problem is how to apply parity conjugates of the $\bar{Q}$ equations, that are the $Q^{(1)}$ equations, to the computation of N$^{k>1}$MHV amplitudes. A good start point would be one- and two-loop N$^{2}$MHV octagons, which are parity invariant by themselves and hence can be fixed by the $\bar{Q}$ equations and the parity symmetry in principle. Similarly, the two-loop N$^{3}$MHV 10-point amplitude can be fixed in the same way, it would be amusing to see how the elliptic polylogarithms arise in this method.

Besides the application in the computations of MHV/NMHV amplitudes, another major application of the $\bar{Q}$ equations is to constrain the last entries and the last two entries~\cite{He:2021mme} for amplitudes. On the one hand, it is worth exploring the relation of such constraints with extended Steinmann/cluster adjacency, on the other hand, similar constraints also appear in the bootstrap of certain form factors~\cite{Brandhuber:2012vm,Dixon:2020bbt}, it would be interesting to generalize the $\bar{Q}$ equations to these cases and investigate other non-perturbative applications of the $\bar{Q}$ equations.

\begin{acknowledgments}
We thank Song He, Andrew McLeod, Matt von Hippel, and Matthias Wilhelm for comments on the manuscript and discussions. 
We would also like to thank Yichao Tang, Cristian Vergu, and Qinglin Yang for discussions. 
ZL is supported in part by National Natural Science Foundation of China under Grant No. 11935013, 11947301, 12047502 and 12047503. 
CZ is supported in part by the ERC starting grant 757978 and grant 00025445 from the Villum Fonden.
\end{acknowledgments}

\appendix

\section{Symbol integration method} \label{appa}

Here we first prove the symbol integration method described in section \ref{sec:2.2}. Consider an integral
\[
\int_a^b {\rm d}\log(t+c)\, (F(t)\otimes w(t)),
\]
where $F(t)\otimes w(t)$ is a linear reducible symbol in $t$, {\it i.e.} its entries are products of powers of linear polynomials in $t$, and $w(t)$ is the last entry. The total differential of this integral is the sum of the following two parts:
\begin{compactenum}[\quad (1)]
\item the contribution from endpoints:
\[
   \Bigl( {\rm d}\log(t+c)F(t)\otimes w(t)\Bigr)\Bigr|_{t=a}^{t=b}=\Bigl(F(t)\otimes w(t)\otimes (t+c)\Bigr)\Bigr|_{t=a}^{t=b},
\]
\item contributions from the last entry: for a term where $w(t)$ is a constant,
\[
\left(\int_a^b {\rm d}\log(t+c)\, F(t)\right){\rm d}\log w=\left(\int_a^b {\rm d}\log(t+c)\, F(t)\right)\otimes w,
\]
and for a term where $w(t)=t+d$,
\[
\left(\int_a^b {\rm d}\log \frac{t+c}{t+d}\, F(t)\right){\rm d}\log (c-d)
=\left(\int_a^b {\rm d}\log \frac{t+c}{t+d}\, F(t)\right)\otimes (c-d).
\]
\end{compactenum}

\begin{proof}
We use $D$ to represent the total differential with respect to variables other than $t$. Therefore, the total differential of the integral is
\begin{align*}
D\int_a^b {\rm d}\log(t+c)\, (F(t)\otimes w(t))
=&\,\frac{Db}{b+c}F(b)\otimes w(b)-\frac{Da}{a+c}F(a)\otimes w(a)\\
&+Dc\int_a^b {\rm d}\biggl(\frac{1}{t+c}\biggr)\, (F(t)\otimes w(t))\\
&+\int_a^b {\rm d}\log(t+c)\, F(t) D\log w(t).
\end{align*}
The second line equals to 
\[
    \frac{Dc}{b+c}F(b)\otimes w(b)-\frac{Dc}{a+c}F(a)\otimes w(a)-Dc\int_a^b \dif\log(t+c)\, F(t) \partial_t \log w(t)
\]
by integrating by parts, so
\begin{align*}
    D\int_a^b {\rm d}\log(t+c)\, (F(t)\otimes w(t))&=\frac{D(b+c)}{b+c}F(b)\otimes w(b)-\frac{D(a+c)}{a+c}F(a)\otimes w(a)\\
    &\quad +\int_a^b {\rm d}\log(t+c)\, (D-Dc\,\partial_t)(\log w(t))\,F(t).
\end{align*}
The first line is exactly the contribution from the endpoints, and the second line is the contribution from the last entry.

Now suppose that the last entry $w(t)=pt+q$ is a linear function of $t$, then 
\begin{align*}
    {\rm d}\log(t+c)\, (D-Dc\,\partial_t)&(\log w(t))=\frac{\dif t}{t+c}\frac{1}{pt+q}\biggl(t Dp+Dq-p Dc\biggr)\\
    &=-D\log(q/p - c)\,\dif \log (pt+q)+D\log (cp-q)\, \dif \log (t+c)
\end{align*}
by a partial fraction which completes the proof.
\end{proof}

There is an interesting subtle for the above method:
\[
\int_a^b {\rm d}\log(t+c)\, F(t),
\]
is not invariant under the variable substitution in general if $F(t)$ is not integrable. To show it, we consider two simpler integrals $\int_a^b \dif\log(t) F(t)$ and $\int_{ac}^{bc} \dif\log(t) F(t/c)$ for a non-integrable symbol $F(t)$. It's clear that we make a variable substitution $t\mapsto ct$. If $F$ is integrable, these two integrals should be the same and independent of $c$, but this is not true for a non-integrable symbol $F(t)$. In fact, the partial differential with respect to $c$ of the last integral is 
\begin{align*}
    c\partial_c\int_{ac}^{bc} \dif\log(t) F(t/c) & = F(b)-F(a)+\int_{ac}^{bc}\dif\log(t) \, c\partial_cF(t/c)\\
    &= F(b)-F(a)-\int_{ac}^{bc} \dif\log(t)\,  t\partial_tF(t/c).
\end{align*}
This may not vanish, \textit{i.e.} results of two integrals by applying the symbol integration method are different. For example, $F(t)=(t+d)\otimes t$, then
\[
    t\partial_tF(t/c)=\log(t/c+d) t\partial_t\log(t/c)=\log(t/c+d)
\]
and 
\[
    \int_{ac}^{bc} \dif\log(t)\,  t\partial_tF(t/c)=F(b)-F(a)+\frac{b (a+d)}{a (b+d)}\otimes d.
\]
In other words, $\int_a^b \dif F(t)$ usually doesn't only depend on two end points $F(b)-F(a)$ if $F$ is not integrable. 

This fact will bother us when the symbol involves irrational objects: the integrable integrand decomposes into two parts
\begin{equation}
    I=\int_a^b \dif\log (t+c) \: \bigl(F_1(t)+F_2(t)\bigr),
\end{equation}
where $F_1$ is purely rational and $F_2$ is not, and they are \emph{not} integrable individually. The obstruction of applying the symbol integration is that some letters of $F_2$ are not rational in $t$. However, once we find a variable substitution $t\mapsto t(z)$ such that letters of $F_2(z)$ and $t+c$ can be written as products of linear factors of $z$, we want to apply the symbol integration procedure on 
\[
    I_1=\int_a^b \dif\log (t+c) F_1(t)\quad \text{and}\quad I_2=\int_{z(a)}^{z(b)} \dif\log (t(z)+c) F_2(z)
\]
individually, then add them together to get the answer of the original integral. Unfortunately, this may lead to a wrong answer. The reason is very simple, suppose that $t+c$ is written as products of linear factors of $z$, \textit{i.e.}
\[
    t(z)+c=c_0\prod_i (z+a_i)^{n_i},
\]
the partial differential with respect to $c_0$ of $I_2=\int_a^b \dif \log(t+c)F_2(t)$
may not vanish if $F_2$ is not integrable, but we miss it in the integral 
\[
    \sum_{i} n_i\int_{z(a)}^{z(b)} \dif\log (z+a_i) F_2(z).
\]

The solution to this problem is also simple: just adding the missing part back
\[
    I_{c_0}:=\int_a^b \dif\log (t) F_1(t)-\int_{c_0a}^{c_0b} \dif\log (t) F_1(t/c_0)\,\,\text{``=''} \int \dif\log(t/(c_0t))F_1(t),
\]
which corresponds to the partial differential with respect to $c_0$ in the total differential, then the original integral equals to 
\[
    \int_a^b \dif\log \frac{t+c}{c_0} (F_1(t)+F_2(t))=I_{c_0}+\int_a^b \dif\log (t+c) F_1(t)+ \sum_{i} n_i\int_{z(a)}^{z(b)} \dif\log (z+a_i) F_2(z),
\]
where we divide a $c_0$ in the $\dif \log$ by hand, but it does not affect the result since $F_1+F_2$ is integrable. Each ``integral'' on the right-hand side of the above equation can be carried out by the symbol integration method described above, and it's also direct to prove that acting with $\partial_{c_0}$ on the right-hand side gives zero.

\section{One-fold integrals with a quadratic curve}

As reviewed in section \ref{sec:2}, there is an automated algorithm for computing the symbol of one-fold integrals of the form
\begin{equation}
    \int_{x_{i}}^{x_{f}} \dif \log(x+\alpha) \: F_{n}(x,\{\beta_{j}\}) \:,  \label{A.1}
\end{equation}
where $F_{n}$ is a weight $n$ polylogarithm whose letters are rational in $x$. In the computation of multi-loop octagons and beyond \cite{Zhang:2019vnm,He:2020vob} and certain Feynman integrals~\cite{He:2020lcu}, we will encounter one-fold integrals of the form
\begin{equation}
    \mathscr{I}_{n+1}=\int_{x_{i}}^{x_{f}} \dif \log r(x,y) \: G_{n}(x,y,\{\beta_{j}\})  \label{A.2}
\end{equation}
where $r(x,y)$ is some rational function in $x,y$ and $G_{n}$ is a weight $n$ polylogarithm whose letters are rational in $x,y$ with a quadratic curve \[
    y^{2}=x^{2}-2ax+b \:.
\] 
These integrals of the form \eqref{A.2} can be reduced to the cases in \eqref{A.1} through a variable substitution, say 
\begin{equation}
    x=\frac{t^{2}-b/4}{t-a/2}\:.  \label{A.3}
\end{equation}
Sometimes, an algorithm ``without'' variable substitutions is more convenient for computing the symbol of the integrals of the form \eqref{A.2}. In this appendix, we will provide such an algorithm by taking turns to use the variable substitution \eqref{A.3} and the algorithm \eqref{symbolint1} -- \eqref{symbolint3}.

The key point is that $\dif \log r(x,y)$ can be expanded in the basis
\begin{equation}
    \mathbb{Q}\biggl[\frac{\dif x}{x-\alpha},\frac{\dif x}{y},\frac{y_{\beta}\dif x}{y(x-\beta)}\biggr] \:,
\end{equation}
where $y_{\beta}=\sqrt{\beta^{2}-2a\beta+b}$. Indeed, one can check that
\begin{align}
    \frac{\dif x}{y}=\frac{\dif t}{t-a/2}\:, \qquad \frac{y_{\beta}\dif x}{y(x-\beta)}= \frac{\dif t}{t-Y_{\beta}^{+}/2}- \frac{\dif t}{t-Y_{\beta}^{-}/2} \label{dxtodt}
\end{align}
are of $\dif \log$-form in $t$, where $Y_{\beta}^{\pm}=\beta\pm y_{\beta}$, and the other terms, like $(x-a)^{-2}y^{-1}$, in $\dif \log r(x,y)=(r_{1}(x)+r_{2}(x)/y) \dif x$ are forbidden since they have singularities other than logarithmic singularities.

In general, the total differential of $G_{n}$ in \eqref{A.2} can be written as
\begin{equation}
    \dif G_{n} = \sum G_{n-1}^{(i)} \dif \log p_{i}(x,y),
\end{equation}
where $p_{i}(x,y)$ is a polynomial in $x$ and $y$. Then, the total differential of $\mathscr{I}_{n+1}$ receives contributions from:
\begin{enumerate}[label=(\roman*)]
     \item The boundary term
    \begin{equation}
         \dif \log r(x_{f},y_{x_{f}}) G_{n}(x_{f},y_{x_{f}})- \dif \log r(x_{i},y_{x_{i}}) G_{n}(x_{i},y_{x_{i}}) \:. 
     \end{equation}
     \item Suppose that the coefficient of the leading term in the expansion of $p_{i}(x,y)$ around $x=\infty$ is $c_{i}$, then there is
     \begin{equation}
         \sum \dif \log c_{i} \int G_{n-1}^{(i)}\: \dif \log r(x,y) \:. 
     \end{equation}
     \item Assume that the $\log r(x,y)$ and $\log p_{i}(x,y)$ have the following expansions
     \begin{align*}
         \frac{\partial\log r(x,y)}{\partial x}\,\dif x&= \sum_{\mu} \frac{A_{\mu}\dif x}{x-\alpha_{\mu}} + \sum_{\nu} \frac{B_{\nu}\, y_{\beta_{\nu}}\dif x}{y(x-\beta_{\nu})} + \frac{C\,\dif x}{y} \:, \\
         \frac{\partial\log p_{i}(x,y)}{\partial x}\,\dif x&= \sum_{\rho} \frac{\tilde{A}_{\rho}\,\dif x}{x-\tilde{\alpha}_{\rho}} + \sum_{\sigma} \frac{\tilde{B}_{\sigma}\, y_{\tilde{\beta}_{\sigma}}\dif x}{y(x-\tilde{\beta}_{\sigma})} + \frac{\tilde{C}\,\dif x}{y}  \:.
     \end{align*}
     Then there is 
     \begingroup
\allowdisplaybreaks
     \begin{align}
         &\quad  \sum_{\mu,\rho}A_{\mu}\tilde{A}_{\rho} \dif \log(\alpha_{\mu}-\tilde{\alpha}_{\rho})\biggl(\frac{\dif x}{x-\alpha_{\mu}}-\frac{\dif x}{x-\tilde{\alpha}_{\rho}}\biggr) \nonumber  \\
         &+\sum_{\mu,\sigma}\frac{A_{\mu}\tilde{B}_{\sigma}}{2}\Biggl( \dif \log\frac{a-Y_{\tilde{\beta}_{\sigma}}^{+}}{a-Y_{\tilde{\beta}_{\sigma}}^{-}} \biggl(\frac{\dif x}{x-\alpha_{\mu}}-\frac{\dif x}{y}\biggr) - 2\,\dif \log(\alpha_{\mu}-\tilde{\beta}_{\sigma}) \frac{y_{\tilde{\beta}_{\sigma}}\,\dif x}{y(x-\tilde{\beta}_{\sigma})} \nonumber \\
         &\qquad  \qquad  \qquad  +\dif \log\frac{\bigl(Y^{-}_{\tilde{\beta}_{\sigma}}-Y^{-}_{\alpha_{\mu}}\bigr)\bigl(Y^{+}_{\tilde{\beta}_{\sigma}}-Y^{+}_{\alpha_{\mu}}\bigr)}{\bigl(Y^{+}_{\tilde{\beta}_{\sigma}}-Y^{-}_{\alpha_{\mu}}\bigr)\bigl(Y^{-}_{\tilde{\beta}_{\sigma}}-Y^{+}_{\alpha_{\mu}}\bigr)} \frac{y_{\alpha_{\mu}}\dif x}{y(x-\alpha_{\mu})}\Biggr) \nonumber \\
         &+\sum_{\mu} \frac{A_{\mu}\tilde{C}}{2} \Biggl( \dif \log\frac{a^{2}-b}{4} \biggl(\frac{\dif x}{x-\alpha_{\mu}}-\frac{\dif x}{y}\biggr) 
          +\dif \log\frac{a-Y^{+}_{\alpha_{\mu}}}{a-Y^{-}_{\alpha_{\mu}}}\, \frac{y_{\alpha_{\mu}}\,\dif x}{y(x-\alpha_{\mu})} \Biggr) \nonumber \\
         &+\sum_{\nu,\rho} \frac{B_{\nu}\tilde{A}_{\rho}}{2}\Biggl( 2\,\dif \log(\tilde{\alpha}_{\rho}-\beta_{\nu})\,\frac{y_{\beta_{\nu}}\,\dif x}{y(x-\beta_{\nu})} 
          +\dif \log\frac{a-Y^{+}_{\beta_{\nu}}}{a-Y^{-}_{\beta_{\nu}}} \biggl(\frac{\dif x}{y}-\frac{\dif x}{x-\tilde{\alpha}_{\rho}}\biggr)  \nonumber \\
         & \qquad  \qquad  \qquad  + \dif \log\frac{\bigl(Y_{\tilde{\alpha}_{\rho}}^{+}-Y_{\beta_{\sigma}}^{-}\bigr)\bigl(Y_{\tilde{\alpha}_{\rho}}^{-}-Y_{\beta_{\sigma}}^{+}\bigr)}{\bigl(Y_{\tilde{\alpha}_{\rho}}^{-}-Y_{\beta_{\sigma}}^{-}\bigr)\bigl(Y_{\tilde{\alpha}_{\rho}}^{+}-Y_{\beta_{\sigma}}^{+}\bigr)} \,\frac{y_{\tilde{\alpha}_{\rho}}\,\dif x}{y(x-\tilde{\alpha}_{\rho})}\Biggr)  \nonumber   \\ 
         &+\sum_{\nu,\sigma} \frac{B_{\nu}\tilde{B}_{\sigma}}{2}\Biggl( \dif \log\frac{a-Y_{\tilde{\beta}_{\sigma}}^{+}}{a-Y_{\tilde{\beta}_{\sigma}}^{-}}\frac{y_{\beta_{\nu}}\,\dif x}{y(x-\beta_{\nu})} -\dif \log \frac{a-Y_{\beta_{\nu}}^{+}}{a-Y_{\beta_{\nu}}^{-}}  \frac{y_{\tilde{\beta}_{\sigma}}\,\dif x}{y(x-\tilde{\beta}_{\sigma})} \nonumber  \\ 
         &\qquad  \qquad  \qquad  +\dif \log \frac{\bigl(Y_{\beta_{\nu}}^{-}-Y_{\tilde{\beta}_{\sigma}}^{-}\bigr)\bigl(Y_{\beta_{\nu}}^{+}-Y_{\tilde{\beta}_{\sigma}}^{+}\bigr)}{\bigl(Y_{\beta_{\nu}}^{-}-Y_{\tilde{\beta}_{\sigma}}^{+}\bigr)\bigl(Y_{\beta_{\nu}}^{+}-Y_{\tilde{\beta}_{\sigma}}^{-}\bigr)} \biggl(\frac{\dif x}{x-\beta_{\nu}}- \frac{\dif x}{x-\tilde{\beta}_{\sigma}}\biggr)\Biggr)  \nonumber  \\
         &+\sum_{\nu} \frac{B_{\nu}\tilde{C}}{2} \Biggl( \dif \log\frac{a-Y_{\beta_{\nu}}^{+}}{a-Y_{\beta_{\nu}}^{-}} \biggl(\frac{\dif x}{x-\beta_{nu}}-\frac{\dif x}{y}\biggr) + \dif \log\frac{a^{2}-b}{4} \frac{y_{\beta_{\nu}}\,\dif x}{y(x-\beta_{\nu})} \Biggr) \nonumber \\
         &+\sum_{\rho}\frac{C\tilde{A}_{\rho}}{2} \Biggl( \dif \log\frac{a^{2}-b}{4}\biggl(\frac{\dif x}{y}-\frac{\dif x}{x-\tilde{\alpha}_{\rho}}\biggr) 
          - \dif \log\frac{a-Y_{\tilde{\alpha}_{\rho}}^{+}}{a-Y_{\tilde{\alpha}_{\rho}}^{-}}  \,\frac{y_{\tilde{\alpha}_{\rho}}\,\dif x}{y(x-\tilde{\alpha}_{\rho})} \Biggr) \nonumber     \\
          &+\sum_{\sigma}\frac{C\tilde{B}_{\sigma}}{2} \Biggl( \dif \log\frac{a-Y_{\tilde{\beta}_{\sigma}}^{+}}{a-Y_{\tilde{\beta}_{\sigma}}^{-}}\biggl(\frac{\dif x}{y}-\frac{\dif x}{x-\tilde{\beta}_{\sigma}}\biggr) 
          - \dif \log\frac{a^{2}-b}{4}  \,\frac{y_{\tilde{\beta}_{\sigma}}\,\dif x}{y(x-\tilde{\beta}_{\sigma})} \Biggr) \:,
     \end{align}
     \endgroup
        where we have omitted $\int G^{(i)}_{n-1}$ for saving space.
    \end{enumerate}
   The above formula is nothing but a result of using the transformation \eqref{dxtodt}, the algorithm \eqref{symbolint1} -- \eqref{symbolint3}, as well as the inverse of \eqref{dxtodt} in turn.

\bibliography{3loopocta} 
\bibliographystyle{JHEP}

\end{document}